\documentstyle[aps,floats,epsf,twocolumn]{revtex}
\begin{document} \draft

\title{Theory of equilibrium flux lattices in unconventional superconductors}

\author{M. Franz$^{1}$, Ian Affleck$^{2,3}$
and M. H. S. Amin$^2$}
\address{$^1$Department of Physics and Astronomy, Johns Hopkins University,
Baltimore, MD 21218\\
$^2$Department of Physics and Astronomy
and $^3$Canadian Institute for Advanced
Research,\\ University of British Columbia, Vancouver, BC, V6T
1Z1, Canada
\\ {\rm(\today)}}
%
\address{~
\parbox{14cm}{\rm
\medskip
We investigate equilibrium flux lattice structures in superconductors
with unconventional order parameters, such as  high-$T_c$ 
cuprates, using a generalized London model with non-local electrodynamics
derived from a simple microscopic model. We find a rich phase diagram
containing triangular, centered rectangular and square lattices with
various orientations relative to the ionic lattice, as a function of 
magnetic field and temperature.
}}
\maketitle


%
\narrowtext
Existence of a mixed state, characterized by a regular array of 
magnetic flux lines penetrating the material, is perhaps one of the most
striking properties of type II superconductors. The original pioneering work
of Abrikosov\cite{aaa}, based on the
 solution of Ginzburg-Landau (GL) equations 
near the upper critical field $H_{c2}$, predicted a triangular flux lattice
(FL). This 
prediction was subsequently verified by low field magnetic decoration 
experiments on a variety of conventional superconductors. In some compounds
neutron scattering experiments revealed 
deviations from perfect triangular lattices in stronger fields 
\cite{review1} which where attributed to anisotropies in the electronic
band structure and other effects and were modeled by GL theories containing 
additional higher order derivative terms reflecting the
material anisotropies\cite{Takanaka}. 

One would expect 
even richer behavior of flux lattices in the new class of heavy fermion and 
copper-oxide superconductors as these exhibit highly anisotropic
electronic structures and, very likely, order parameters with 
unconventional symmetries involving nodes in the gap. 
In high-$T_c$ cuprates much of the experimental and 
theoretical effort has been  focused on the sizable region of the phase 
diagram just below $T_c(H)$ in which the vortex lattice properties are 
dominated
by thermal fluctuations\cite{review3}. While understanding the physics of this 
fluctuation dominated regime poses an intriguing and 
difficult statistical mechanics
problem, investigation of the equilibrium vortex lattice 
structures at low temperatures may provide clues about the microscopic 
mechanism in these materials. So far such investigations
have been limited to YBa$_2$Cu$_3$O$_{7-\delta}$  compound 
\cite{Keimer,Maggio} , revealing
vortex lattices with centered rectangular symmetry and various
orientations relative to the ionic lattice. These have been modeled by
phenomenological GL theories appropriate for anisotropic superconductors,
containing additional quartic derivative terms
\cite{quartic} or a mixed gradient coupling to an order parameter with
different symmetry\cite{sd}. These works found 
structures in qualitative agreement with experiment, but their inherent 
shortcoming is the large number of unknown phenomenological parameters and 
the subsequent lack of predictive power. Also, such GL theory has only 
been solved
for vortex lattice near $H_{c2}$, which is experimentally inaccessible in
cuprates away from $T_c$. [See however Ref.\cite{brandt2} which holds
promise for full solution at any field.] 
We have recently formulated a generalized London 
model\cite{Affleck} which is valid 
in experimentally accessible region of intermediate fields 
$H_{c1}<H\ll H_{c2}$. This model is also phenomenological and it 
contains one unknown parameter which
controls the strength of the symmetry breaking term.
With increasing magnetic field this
model predicts a transition from triangular to 
square FL. While no direct experimental evidence
exists in cuprates at present to confirm such a prediction, a similar 
transition has been recently observed in a boro-carbide material ErNi$_2$B$_2$C
\cite{Yaron} and has been described by a similar London model\cite{Kogan1}.

In this letter we present a microscopic derivation of the generalized 
non-local London model for an unconventional superconductor.
Based on this model we formulate, for the first time,
{\em quantitative} and largely {\em parameter free} predictions for the 
behavior of the vortex lattice structure
as a function of temperature and magnetic field. Our theory is valid in a
large part of the $H$-$T$ phase diagram, only restricted by the inherent
domain of validity of the London model, $H\ll H_{c2}$, and $T$ low enough
that the thermal fluctuations are unimportant. 
The central result of this work is a prediction 
that the FL geometry in unconventional
superconductors will display a rich and distinctive behavior as a function
of field and temperature, undergoing a series of transitions and crossovers, 
and eventually attaining a {\em universal} limit at
$T=0$. The London free energy in this limit is non-analytic and its long 
wavelength part is fully determined by the nodal structure of the gap 
function. Such behavior is caused by the low-lying quasiparticle excitations 
within the nodes and thus could never occur in conventional superconductors
with anisotropic band structures.

In general
the relation between the supercurrent ${\bf j}$ and the vector potential
${\bf A}$ of the magnetic induction ${\bf B}=\nabla\times{\bf A}$ is 
non-local in a superconductor, 
reflecting the finite spatial extent of the Cooper pair 
$\sim\xi_0$\cite{Tinkham}. 
Non-local corrections to physical quantities, such as the effective
penetration depth, will be of order $\kappa^{-2}$, where 
$\kappa\equiv \lambda_0/\xi_0$ is the GL ratio and $\lambda_0$
is the London penetration depth. For strongly type 
II materials ($\kappa\gg 1$) such corrections are 
negligible. Since cuprate superconductors fall well within this class 
($\kappa$ is in excess of 50 for most) local 
electrodynamics is always used. However, a closer examination suggests that 
this might not be justified in all situations, if, as it is widely believed,
these materials exhibit nodes in the gap. In such a case 
in place of the usual coherence length $\xi_0=v_F/\pi\Delta_0$
one is forced to define an angle dependent quantity, $\xi_0({\bf\hat p})=
v_F/\pi\Delta_{\bf\hat p}$, which diverges along the nodes. Clearly, in the
vicinity of nodes the condition  $\lambda_0/\xi_0({\bf\hat p})\gg 1$ is no 
longer 
satisfied and, in fact, the extreme {\em non-local} limit is achieved.
Non-local corrections therefore cannot be dismissed in unconventional
superconductors, especially at low temperatures when quasiparticles 
selectively populate nodal regions.
 From the above argument it is also clear that such corrections
will be highly anisotropic and will in general break the rotational symmetry
of the flow field around the vortex, contributing an anisotropic
component to the inter-vortex interaction in the mixed state. Thus, on very
general grounds, one may 
expect non-triangular FL structures in unconventional superconductors.
We now illustrate this idea by computing the FL geometry for the simplest
case of a $d_{x^2-y^2}$ superconductor with cylindrical fermi surface in the 
clean limit. 
 
The non-local relation between ${\bf j}$ and ${\bf A}$ is conveniently written
in Fourier space\cite{Tinkham}
\begin{equation}
{\bf j_k}=-(c/4\pi){\bf \hat Q(k) A_k}.
\label{n1}
\end{equation}
Here ${\bf \hat Q(k)}$ is the electromagnetic response tensor which can be 
computed, within the weak coupling theory, by generalizing the standard 
linear response treatment of Gorkov equations\cite{Lif} to an anisotropic gap. 
We find
\begin{equation}
Q_{ij}({\bf k})={4\pi T\over \lambda_0^2}\sum_{n>0}\left\langle
{\Delta_{\bf\hat p}^2\hat v_{Fi}\hat v_{Fj} \over 
\sqrt{\omega_n^2+\Delta_{\bf\hat p}^2}
(\omega_n^2+\Delta_{\bf\hat p}^2+\gamma_{\bf k}^2)}\right\rangle,
\label{Q}
\end{equation}
where $\gamma_{\bf k}={\bf v}_F\cdot{\bf k}/2$, London penetration depth
$\lambda_0^{-2}=4\pi e^2v_F^2N(0)/c^2$, Matsubara frequencies
$\omega_n=\pi T(2n-1)$ and the angular brackets mean the Fermi surface 
averaging. Eq.(\ref{Q}) is valid for arbitrary Fermi surface and gap function.
For isotropic gap one recovers an expression recently derived
by Kogan {\em et al.\ }from the Eilenberger theory\cite{Kogan2}.
One may simplify solving the London equation by writing it 
in terms of magnetic induction only.
Eliminating ${\bf j}$ from  Eq.(\ref{n1}) using the 
Amp\`{e}re's law ${\bf j}=(c/4\pi)\nabla\times{\bf B}$, one obtains
\begin{equation}
{\bf B_k}-{\bf k}\times[{\bf \hat Q^{-1}(k)}({\bf k}\times{\bf B_k})]=0. 
\label{london1}
\end{equation}
For many purposes it is also convenient to write down the corresponding
London free energy $F_L$, such that $\delta F_L/\delta {\bf B_k}=0$ 
gives the above London equation:
\begin{equation}
F_L=\sum_{\bf_k}[{\bf B_k}^2+ ({\bf k}\times{\bf B_k})
{\bf \hat Q^{-1}(k)}({\bf k}\times{\bf B_k})]/8\pi.
\label{free}
\end{equation}
It is easy to see that in the local limit 
$Q_{ij}(k\to 0)=\delta_{ij}\lambda^{-2}$ one recovers the usual London
free energy\cite{Tinkham}.
Here $\lambda\equiv\lambda(T)$ is the temperature dependent penetration 
depth (given below) for which it holds that $\lambda(0)=\lambda_0$.

One may study FL structure using this formalism provided the cores
occupy only
a small fraction of the total volume, i.e. when $H\ll H_{c2}$ and 
$\kappa\gg 1$. To account for the topological winding of the phase around 
the core\cite{Affleck,Tinkham} it is then necessary to insert source
terms $\rho_{\bf k}$ on the right-hand side of Eq.(\ref{london1}). 
A commonly used from is\cite{Brandt}
\begin{equation}
\rho_{\bf k}=(\phi_0/\Omega)e^{-k^2\xi^2/2}, 
\label{source}
\end{equation}
where the prefactor insures proper flux quantization 
($\phi_0$ is the flux quantum and $\Omega$ is the area of the FL unit cell).
For a given kernel ${\bf \hat Q(k)}$ the FL symmetry is then determined
by minimizing the Gibbs free energy $G_L=F_L- H\bar B/4\pi$ 
(where $\bar B=\phi_0/\Omega$ is the average induction).

At long wavelengths ${\bf \hat Q(k)}$ can be evaluated by expanding expression
(\ref{Q}) in powers of $\gamma_{\bf k}^2$. The zeroth order term 
\begin{equation}
Q_{ij}^{(0)}\equiv\delta_{ij}\lambda^{-2}
={4\pi T\over \lambda_0^2}\sum_{n>0}\left\langle
{\Delta_{\bf\hat p}^2\hat v_{Fi}\hat v_{Fj} \over 
(\omega_n^2+\Delta_{\bf\hat p}^2)^{3/2}}\right\rangle,
\label{Q0}
\end{equation}
is just the temperature dependent penetration depth which, at low 
temperatures, has the well known $T$-linear behavior\cite{Scalapino}:
$\lambda^{-2}\approx\lambda_0^{-2}[1-(2\ln2)T/\Delta_d]$ for a $d_{x^2-y^2}$
superconductor with $\Delta_{\bf\hat p}=\Delta_d(\hat p_x^2-\hat p_y^2)$
and a cylindrical Fermi surface. From now on we shall focus on this 
simple case. The leading non-local term is quadratic in $k$:
\begin{equation}
Q_{ij}^{(2)}
=-{4\pi T\over \lambda_0^2}\sum_{n>0}\left\langle
{\Delta_{\bf\hat p}^2\hat v_{Fi}\hat v_{Fj} \over 
(\omega_n^2+\Delta_{\bf\hat p}^2)^{5/2}}\gamma_{\bf k}^2 \right\rangle.
\label{Q2}
\end{equation}
The expression $Q_{ij}= \delta_{ij}\lambda^{-2}+Q_{ij}^{(2)}$ is
easily inverted to leading order in $k$:  
$Q_{ij}^{-1}\approx\lambda^2[\delta_{ij}-\lambda^2 Q_{ij}^{(2)}]$. 
Substituting
this into Eq.(\ref{free}) and specializing to fields along the $z$-direction
we have
\begin{equation}
F_L=\sum_{\bf_k}B_{\bf k}^2[1+ \lambda^2k^2
+\lambda^2\xi^2(c_1k^4+c_2k_x^2k_y^2)]/8\pi.
\label{free-anal}
\end{equation}
Here $\xi=v_F/\pi\Delta_d$ and $\Delta_d$ is assumed to be a temperature
dependent solution to the appropriate gap equation. Dimensionless coefficients
$c_1$ and $c_2$ are given by
\renewcommand{\arraystretch}{1}
\begin{equation}
c_\mu
={\lambda^2\over\lambda_0^2}\pi^3\Delta_d^2T
\sum_{n>0}{1\over 2\pi}\int_0^{2\pi}d\theta
{\Delta_{\bf\hat p}^2 w_\mu \over (\omega_n^2+\Delta_{\bf\hat p}^2)^{5/2}},
\label{c12}
\end{equation}
where $w_1=\hat v_{Fx}^2 \hat v_{Fy}^2$, 
$w_2=(\hat v_{Fx}^2 -\hat v_{Fy}^2)^2-4\hat v_{Fx}^2 \hat v_{Fy}^2$ and 
the Fermi surface has been explicitly parameterized by the angle $\theta$
between ${\bf\hat p}$ and the 
$x$ axis: ${\bf\hat v}_F=(\cos\theta,\sin\theta)$ and
$\Delta_{\bf\hat p}=\Delta_d\cos2\theta$. 
Coefficients $c_1$ and $c_2$ depend on temperature through a dimensionless
parameter $t\equiv T/\Delta_d$. From Eq.(\ref{c12}) one can deduce their
leading behavior in the two limiting cases: for $t\ll 1$ we find
\begin{equation}
c_1={\pi^2\over 8}{\lambda^2\over\lambda_0^2}{1\over t}, \ \ c_2=-4c_1,
\label{clow}
\end{equation}
and for $t\gg 1$ (i.e., near $T_c$)
\begin{equation}
c_1=\alpha{\lambda^2\over\lambda_0^2}{1\over t^4}, \ \ c_2 =8c_1,
\label{chigh}
\end{equation}
where $\alpha=\zeta(5)(1-2^{-5})/8\pi^2=0.01272$. In the above $\lambda$ 
also depends on $t$, but this will be unimportant for the following 
qualitative discussion.

\begin{figure}[t]
\epsfxsize=7.5cm
\epsffile{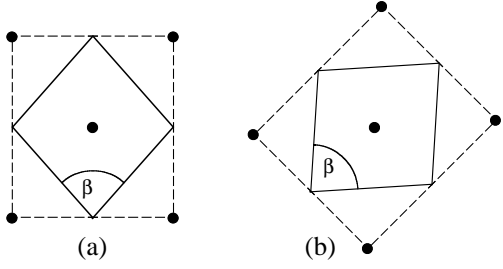}
\caption[]{Two high symmetry orientations of centered rectangular unit cell.
}
\label{fig1}
\end{figure}
The free energy (\ref{free-anal}) formally 
coincides with the one deduced previously from 
phenomenological considerations\cite{Affleck}. 
An interesting new feature emerging from 
 the present microscopic model is a sign
reversal of $c_2$ at intermediate $t$ implied by Eqs. (\ref{chigh}) and
(\ref{clow}).
At high temperatures $c_2$ is found to be positive, in agreement with 
\cite{Affleck}. It has been shown that such a term leads to centered 
rectangular
FL structure with principal axes oriented along $x$ or $y$ axes of the ionic
lattice (cf. Fig.\ \ref{fig1}a). 
The magnitude of the distortion from a perfect triangular lattice 
($\beta=60^\circ$) is 
controlled by the magnitude of $c_2$ and grows with increasing magnetic field.
Eq.(\ref{chigh}) shows that at fixed field 
this distortion will initially grow with decreasing temperature. 
At low temperatures Eq.(\ref{clow}) predicts $c_2<0$. This will
lead to the same centered rectangular FL rotated by $45^\circ$ 
(cf. Fig.\ \ref{fig1}b). Numerical
evaluation of Eq.(\ref{c12}) shows that $c_2$ passes through zero at 
$t^*\simeq0.19$. At this temperature the free energy (\ref{free-anal}) 
is isotropic and the FL will be triangular at all fields. The 
sign reversal of $c_2$ reflects
the competition between the two terms of different symmetry in $w_2$
and is a unique consequence of the gap function having nodes.

Another consequence of nodes
is the fact that, as it can be seen from Eq.(\ref{clow}), both $c_1$ and
$c_2$ diverge as $1/t$ for $t\to 0$.  This divergence signals that the 
response tensor ${\bf\hat Q(k)}$ is a {\em non-analytic} function of ${\bf k}$
at $T=0$ and the expansion in powers of $\gamma_{\bf k}^2$ breaks down. 
Formally this is caused by the fact that at $T=0$ at the nodal point
the expression (\ref{Q}) for $Q_{ij}({\bf k})$ contains a term
proportional to $1/\gamma_{\bf k}^2$. At $T=0$ the frequency sum in (\ref{Q})
becomes an integral which can be evaluated exactly with the result:
\begin{equation}
Q_{ij}({\bf k})={1\over \lambda_0^2}\left\langle
\hat v_{Fi}\hat v_{Fj}
{2{\rm arc\ sinh\ }y \over y\sqrt{1+y^2}} \right\rangle,
\label{Q-nono}
\end{equation}
where $y=\gamma_{\bf k}/\Delta_{\bf\hat p}$. For small $k$ the dominant 
contribution to the angular average comes from the close vicinity of nodes
and can be evaluated by linearizing $\Delta_{\bf\hat p}$ 
around the nodes. One finds that the
leading non-local contribution is {\em linear} in $k$ rather than quadratic.
For $Q_{ij}= \delta_{ij}\lambda_0^{-2}+Q_{ij}^{(1)}$, we have 
$Q_{xx}^{(1)}=Q_{yy}^{(1)}=-\mu(k_>\xi_0)$ and 
$Q_{xy}^{(1)}=Q_{yx}^{(1)}=-\mu(k_<\xi_0)
{\rm sgn}(\hat k_x\hat k_y)$, where $k_>=\max(|k_x|,|k_y|)$ 
and $k_<=\min(|k_x|,|k_y|)$. Prefactor 
$\mu=\pi^2/8\sqrt{2}=0.8723$ is exact in the sense that all corrections to
$Q_{ij}$ are $O(k^2)$. The resulting free energy at $T=0$ is 
\begin{equation}
F_L=\sum_{\bf_k}B_{\bf k}^2[1+ \lambda_0^2k^2
+\mu(\lambda_0^2\xi_0) k_>
(k_>^2- k_<^2)]/8\pi.
\label{free-nono}
\end{equation}
The non-local term is clearly non-analytic in $k$. Its functional form is 
universal in the sense that it is independent of the Fermi surface structure
(as long as it has tetragonal symmetry) and
the prefactor $\mu$ only depends on the angular slope of the gap function and
Fermi velocity at the node. 
Numerical evaluation shows that the free energy (\ref{free-nono}) gives rise to
a centered rectangular FL structure, aligned with $x$ or $y$ axes, but
now with the apex angle $\beta<60^\circ$, depending {\em non-monotonically}
on the field. This suggests that there will be an additional transition at 
low temperature related to the non-analytic behavior of the response tensor.
\begin{figure}[t]
\epsfxsize=8.5cm
\epsffile{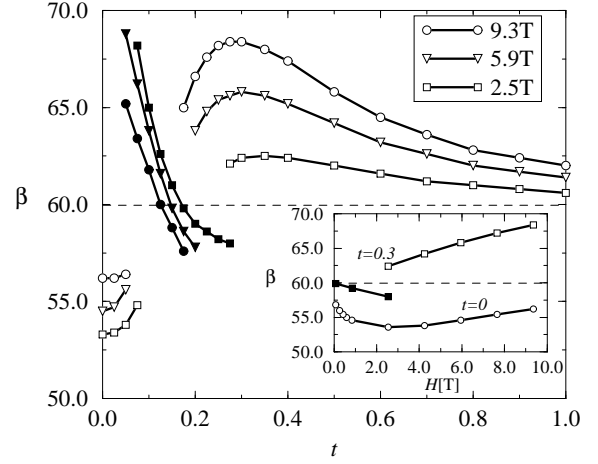}
\caption[]{Equilibrium angle $\beta$ as a function of reduced temperature
$t=T/\Delta_d$ for various fields. Open symbols mark lattice with orientation
along $x$ or $y$ direction while solid symbols mark the lattice rotated
by $45^\circ$. We use $\lambda_0=1400$\AA$$  and $\kappa=68$. Inset: $\beta$ as
 a function of field at fixed $T$.}
\label{fig2}
\end{figure}

In order to map out the complete equilibrium $H$-$T$ phase diagram 
we have carried out a
numerical computation of the FL structure using the full expression 
for the response tensor ${\bf\hat Q(k)}$, as given by Eqs.\ (\ref{Q}) and
(\ref{Q-nono}). We find that free energy has two local minima
for centered rectangular lattices aligned with two high symmetry directions
shown in Fig.\ \ref{fig1}, 
as expected from the tetragonal symmetry of the problem.
Which of the two becomes the global minimum depends on temperature and field.
The results are summarized in Fig.\ \ref{fig2}. For high temperatures the exact
result agrees well with the one obtained from the long wavelength free
energy (\ref{free-anal}). The deformation of the lattice from perfect
triangular grows with decreasing temperature, reaches a maximum, and then 
falls. Maximum distortion occurs around $t\simeq 0.3$, attaining 
$\beta\simeq 70^\circ$ at 10T. Extrapolating this field dependence (see inset
to Fig.\ \ref{fig2}), the FL should become square around $H\approx 30$T, but
this field is outside the domain of validity of the London model.  
At lower temperatures the distortion decreases but instead of 
going all the
way back to triangular at $t^*$, the lattice  undergoes a first order
phase transition to another centered rectangular lattice rotated by $45^\circ$
and with $\beta<60^\circ$. Further decrease of temperature causes the angle
to grow again. We note that precise temperature at which it crosses 
$60^\circ$ depends on field, but for all fields is close to $t^*=0.19$, as 
predicted by the long wavelength approximation. 
At yet lower temperature we predict another first order transition to a 
centered rectangular lattice along $x$ (or $y$) with $\beta<60^\circ$.
The free energy difference between the two minima
is very small in the region where the 
$45^\circ$ rotated lattice wins. It is thus
likely that real system
will remain in the metastable state and the experiment
would detect only a smooth crossover from a lattice with $\beta>60^\circ$
to the one with $\beta<60^\circ$.

The present calculation can be easily generalized to treat the effects of 
Fermi surface anisotropy. As 
mentioned above tetragonal anisotropy will not modify the $T\to 0$ universal 
behavior but may lead to quantitative changes at higher temperatures. 
Orthorhombic anisotropy, on the other hand, will modify even the $T\to 0$
limit. We expect that it will, to leading order, merely rescale the coordinate
axes, leading to the same structures as described above stretched by the
appropriate factor\cite{Affleck}. It may further remove the degeneracy between
two equivalent lattices related by $90^\circ$ rotation. 
Another source of anisotropy neglected in our calculation is the non-linear
Meissner effect 
studied by Yip and Sauls \cite{YS}, associated with the shift of quasiparticle
spectrum due to the superflow. Within the quasiclassical picture this
would contribute terms $\sim(|\partial_{x'}B|^3 + |\partial_{y'}B|^3)$ to the 
London free energy at $T=0$, 
where $x'$ and $y'$ are $45^\circ$ rotated coordinates.
Our initial numerical results\cite{mehrdad} indicate that while not  
completely negligible, these terms will not significantly alter the behavior
of FL reported in Fig.\ \ref{fig2}. Also neglected in 
our calculation is the effect of electronic disorder, which will remove
the non-analyticity of ${\bf\hat Q(k)}$ at longest wavelengths, just as
small finite temperature would. 
Since the FL is most sensitive to  ${\bf\hat Q(k)}$ at
finite $k\sim l^{-1}$ ($l$ is the vortex spacing), we expect our predictions
to be robust with respect to weak disorder.

In conclusion we have described distinctive features of the vortex lattice
geometry associated with unconventional pairing and non-local response. 
Near $T_c$ our predictions are consistent with the existing phenomenological
work\cite{quartic,sd,Affleck} while at low $T$ we predict novel effects
intimately related to the nodal structure of the order parameter, which are
not contained in GL-type theories.   
These will occur simultaneously with other unique effects predicted 
previously, such as the $\sim\sqrt{H}$ dependence of specific 
heat\cite{Volovik}. Existing experiments
probing the FL geometry in cuprates\cite{Keimer,Maggio} provide a somewhat 
conflicting picture and their theoretical analysis is complicated by 
the orthorhombic anisotropy and presence of twin boundaries. We
hope that the present theory will encourage more systematic experimental work,
preferably on untwinned or tetragonal materials. 

After this work was completed we learned about a preprint by Kosztin and 
Legett\cite{Kosztin}
 which discusses effects of non-locality on the effective penetration 
depth in $d$-wave superconductors. Where overlap exists their results appear
consistent with ours.

The authors are indebted to A. J. Berlinsky and Z. Te\v{s}\'{a}novi\`{c} for
helpful discussions.  This research was supported by NSERC, the CIAR and NSF
grant DMR-9415549 (M.F).


\end{document}